\documentclass[twocolumn,3p,times]{elsarticle}
\usepackage[latin9]{inputenc}
\usepackage{amsmath}
\usepackage{amssymb}
\usepackage{graphicx}
\usepackage{color}

\usepackage[figuresright]{rotating}

\begin{document}
\begin{frontmatter}




\title{Determination of the Higgs $CP$-mixing angle in the tau decay
  channels\tnoteref{label2}}

\tnotetext[label2]{Talk given by S. Berge at the $37^{th}$
  International Conference on High Energy Physics (ICHEP), July 2014. }


\author{Stefan Berge, Werner Bernreuther and Sebastian Kirchner}

\address{Institut für Theoretische Physik, RWTH Aachen University, 52056 Aachen,
Germany}
\begin{abstract}
The measurement of possible Higgs sector $CP$-violation in the tau
decay channels at the LHC is investigated. A $CP$-violating effect
would manifest itself in these decay modes in characteristic spin-spin
correlations of the tau lepton pairs which can be accessed using the
momenta and impact parameters of the charged tau decay particles.
We examine a $CP$-sensitive observable for a 125 GeV Higgs boson
resonance in the gluon fusion channel at the LHC. Furthermore, we
consider the distribution of this observable for the irreducible Drell-Yan
background. By splitting these events into two categories we obtain
two different distributions which can be used for calibration purposes.
Finally, we estimate the achievable precision of the scalar-pseudo-scalar
mixing angle of the tau decay channel for Run II and the high luminosity
run of the LHC.\\ 
\vspace*{-11cm}
\begin{flushright}
TTK-14-29\\
\end{flushright}
\vspace*{10.cm}
\vspace{-0.6mm}
\end{abstract}
\end{frontmatter}



\section{Introduction}

In 2012, the ATLAS and CMS collaborations \citep{Aad:2012tfa,Chatrchyan:2012ufa}
discovered a new electrically neutral boson $h$ with mass $m_{h}\simeq125$
GeV, called Higgs boson.
So far experimental results on the properties of $h$~\citep{Chatrchyan:2012jja,Aad:2013xqa,Chatrchyan:2013lba,Aad:2013wqa,Chatrchyan:2013mxa,ATLtauconf,Chatrchyan:2014vua,Chatrchyan:2014nva,ATLHprconf} agree  within errors
with the predictions of the Standard Model (SM). In particular, its  couplings to gauge
bosons and quarks and leptons
also match the SM predictions within the experimental and theoretical
uncertainties. The important question concerning the $CP$-nature
of the Higgs boson could not yet be experimentally determined. While
it is excluded that the Higgs boson is a pure $CP$-odd state, a large
$CP$-odd component is still allowed~\citep{Djouadi:2013qya,Bechtle:2014ewa,Brod:2013cka}.

To establish or exclude possible $CP$-violating interactions of $h$,
one should investigate all Higgs boson couplings to SM particles because
an extended model with Higgs sector $CP$-violation can generate different
impacts on the couplings of SM particles to Higgs bosons. Unfortunately,
$CP$-effects for only a few Higgs couplings will directly be testable
at the LHC. $CP$-sensitive observables in high energy scattering
processes can conveniently be constructed in dependence of decay particle
momenta of intermediate Higgs bosons. A promising channel is the recently
established~\citep{ATLtauconf,Chatrchyan:2014nva} $h\to\tau^{+}\tau^{-}$
decay channel because of its large branching fraction and because
$CP$-violating couplings can already be present at the leading order
of the perturbative expansion. In a series of papers \citep{Berge:2008dr,Berge:2008wi,Berge:2011ij,Berge:2014sra}
we have proposed a method to measure the $CP$-property of $h$ using
the momenta and impact parameters of the charged final particles of
the tau decays. Important aspects of our method are that the tau momenta
do not need to be reconstructed and all major tau decay channels can
be used. 

Another method~\citep{Harnik:2013aja} uses the $h\to\tau^{+}\tau^{-}$
decay channel at the LHC with subsequent $\tau\to\rho+\nu_{\tau}$
decay. Furthermore, the proposal~\citep{Dolan:2014upa} investigates
the mixing angle of the Higgs-top quark coupling $\phi_{t}$ in the
$pp\to h+2jet$ channel with subsequent $h\to\tau^{+}\tau^{-}$ decay.
Here the $\tau$ leptons are used to identify the Higgs boson within
the background processes, rather than to extract the mixing angle
$\phi_{\tau}$ with the help of $\tau$ spin correlations. This production
channel can in principle be used to perform the Higgs $CP$-measurements
of $\phi_{t}$ and $\phi_{\tau}$ at the same time. Finally, if very
large luminosities will be available at the LHC, $CP$-violating effects
might be investigated with suitably chosen observables in the $h\to\tau^{+}\tau^{-}+\gamma$
decay channel~\citep{Chen:2014ona}. 

In addition to the Higgs signal distributions, we also investigated
the distribution of our $CP$-sensitive observable for the irreducible
Drell-Yan background $pp\to Z^{*}/\gamma^{*}\to\tau^{+}\tau^{-}$
including NLO QCD corrections~\citep{Berge:2014sra}. We further
suggest a method how to use these background events at the LHC, to
calibrate the measurement of the mixing angle~$\phi_{\tau}$ and
to determine experimental  uncertainties.

\section{Spin correlations in Higgs decay to $\tau^{+}\tau^{-}$}

We parametrize the Lagrangian for the $CP$-violating Higgs to tau
coupling in dependence of an effective $\tau$-Yukawa interaction
strength $g_{\tau}$ and a scalar-pseudo-scalar mixing angle $\phi_{\tau}$:

\begin{equation}
{\cal L}_{Y}=-g_{\tau}\left(\cos\phi_{\tau}\bar{\tau}\tau+\sin\phi_{\tau}\bar{\tau}i\gamma_{5}\tau\right)h\,.\label{YukLa-phi}
\end{equation}
The mixing angle $\phi_{\tau}$ can be accessed using the differential
decay width of the Higgs boson into a pair of $\tau$ leptons. In
the Higgs rest frame the differential decay width can be written in
dependence of the $\tau$-spins for the approximation $\beta_{\tau}=\sqrt{1-4m_{\tau}^{2}/m_{h}^{2}}\approx1$
as \vspace{-1mm}
\begin{align}
d\Gamma_{h\to\tau^{+}\tau^{-}} & \sim1-s_{z}^{-}s_{z}^{+}+\cos\left(2\phi_{\tau}\right)\left({\bf s}_{T}^{\,-}\cdot{\bf s}_{T}^{\,+}\right)\nonumber \\
 & \qquad\quad{}+\sin\left(2\phi_{\tau}\right)\left[\left({\bf s}_{T}^{\,+}\times{\bf s}_{T}^{\,-}\right)\cdot{\bf \hat{k}}^{-}\right]\,.\label{dGamma_s-s+}
\end{align}
Here ${\bf \hat{k}}^{-}$ is the normalized $\tau^{-}$ momentum in
the Higgs rest frame pointing in the positive $z$ direction. Furthermore
we define by ${\bf \hat{s}}^{\mp}$ the normalized spin vectors of
the $\tau^{\mp}$ in their respective $\tau$ rest frames where these
rest frames are obtained from the Higgs rest frame by a rotation-free
Lorentz boost along the ${\bf \tau}^{\pm}$ momenta. Then $s_{z}^{\mp}$
denote the $z$-components of ${\bf \hat{s}}^{\mp}$ and ${\bf s}_{T}^{\,\mp}$
the transverse vectors of ${\bf \hat{s}}^{\mp}$. Eq.~\eqref{dGamma_s-s+}
shows that the longitudinal spin correlation is not sensitive to the
mixing angle $\phi_{\tau}$ and that the scalar product of the third
term is $CP$-even while the triple product in the fourth term is
$CP$-odd. The $CP$-sensitive information is therefore encoded in
the transverse component of the $\tau^{+}$$\tau^{-}$ spin correlation
(with respect to ${\bf \hat{k}}^{-}$ in the Higgs rest frame). If
$\varphi_{s}$ denotes the angle pointing from ${\bf s}_{T}^{\,+}$
to ${\bf s}_{T}^{\,-}$ in a right handed coordinate system, the differential
decay width can be written as 

\vspace{-3mm}

\begin{equation}
d\Gamma_{h\to\tau^{+}\tau^{-}}\sim1-s_{z}^{-}s_{z}^{+}+\left|{\bf s}_{T}^{\,-}\right|\left|{\bf s}_{T}^{\,+}\right|\cos\left(\varphi_{s}-2\phi_{\tau}\right)\,.\label{dGamma_s-s+-1}
\end{equation}
The $\tau$-spin correlations of ${\bf s}_{T}^{\,+}$ and ${\bf s}_{T}^{\,-}$
can be measured by investigating the angular correlations of the $\tau$
decay products. If one considers the direct $\tau^{-}\to\pi^{-}+\nu_{\tau}$
decay, due to angular momentum conservation, the $\pi^{-}$ momentum
points in the $\tau^{-}$ rest frame preferably in the same direction
as the $\tau^{-}$ spin. Correspondingly for the direct $\tau^{+}\to\pi^{+}+\bar{\nu}_{\tau}$
decay, the $\pi^{+}$ momentum points in the $\tau^{+}$ rest frame
preferably in the opposite direction as the $\tau^{-}$ spin. Similar
statements can be derived for the other $\tau$ decay channels~\citep{Kuhn:1995nn}.

Our method~\citep{Berge:2008wi,Berge:2008dr,Berge:2011ij,Berge:2013jra}
to access the $\tau$-spin correlation between ${\bf s}_{T}^{\,+}$
and ${\bf s}_{T}^{\,-}$ is applicable for all major tau decay channels:
the leptonic decay channel $\tau\to l+\nu_{\tau}+\nu_{l}$, with $l=\{e,\mu\}$,
which includes 2 neutrinos for each $\tau$ decay and the hadronic
decay channels $\tau^{-}\to\pi^{-}+\nu_{\tau}$, $\tau^{-}\to\rho^{-}+\nu_{\tau}$
and $\tau^{-}\to a_{1}^{-}+\nu_{\tau}$ with subsequent decay of the
$\rho$-meson into a charged and a neutral pion and the decay of the
$a_{1}$-meson into either one charged pion plus neutral pions or
the decay into 3 charged pions.

The momentum vectors of the charged prongs%
\footnote{$a_{1}^{L,T,\pm}$ denotes the $a_{1}$ meson with subsequent decay
into 3 charged pions with $L\,(T)$ longitudinal (transverse) helicity
states%
} $a^{\pm},a'^{\pm}\in\{e^{\pm},\mu^{\pm},\pi^{\pm},a_{1}^{L,T,\pm}\}$
of the $\tau^{-}\to a^{-}+X$ and $\tau^{+}\to a'^{+}+X$ decays act
as spin analyzers. If $E_{\mp}$ and $\hat{{\bf q}}^{\mp}$ are the
energies and directions of flight of the $a^{\mp}$ in the respective
$\tau$ rest frame, the normalized $\tau$-decay distributions are
of the form
\begin{align}
 & {\Gamma_{a}}^{-1}\mbox{d}\Gamma\left(\tau^{\mp}(\hat{{\bf s}}^{\mp})\to a^{\mp}(q^{\mp})+X\right)\qquad\qquad\qquad\nonumber \\
 & \qquad\qquad=\, n\left(E_{\mp}\right)\left[1\pm b\left(E_{\mp}\right)\,\hat{{\bf s}}^{\mp}\cdot\hat{{\bf q}}^{\mp}\right]dE_{\mp}\frac{d\Omega_{\mp}}{4\pi}\,.\label{eq:dGamma_dEdOmega}
\end{align}
The spectral functions $b\left(E_{\mp}\right)$ and $n\left(E_{\mp}\right)$
are given in \citep{Berge:2011ij}. The function $b(E_{\mp})$ describes
the $\tau$-spin analyzing power of the decay particle $a^{\mp}$
and is maximal for the direct decays to pions, $\tau^{\mp}\to\pi^{\mp}+\nu_{\tau}/\bar{\nu}_{\tau}$,
and for the decay%
\footnote{The $\tau$-spin analyzing power of $a_{1}^{L-}$ and $a_{1}^{T-}$
is $+1$ and $-1$, respectively.%
} $\tau^{\mp}\to a_{1}^{L,T,\mp}+\nu_{\tau}/\bar{\nu}_{\tau}$. For
the other decays, the $\tau$-spin analyzing powers of $l^{\mp}$
and $\pi^{\mp}$ depends on the energy $E_{\mp}$ .

\begin{figure}[t]
\includegraphics[width=8cm]{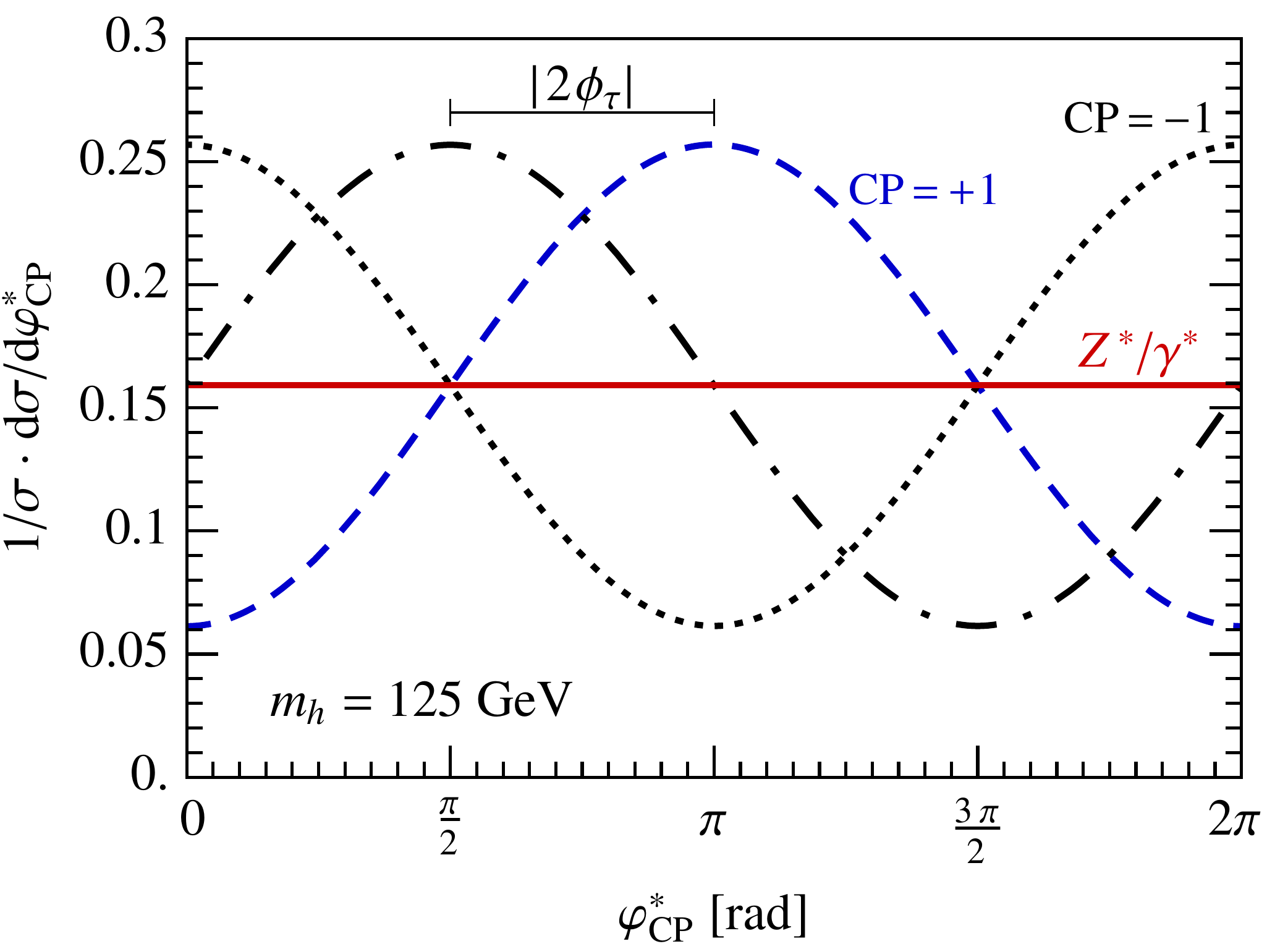}
\caption{Normalized $\varphi_{CP}^{*}$ distribution for $pp\to h/Z^{*}/\gamma^{*}\to\tau^{+}\tau^{-}$
production at the LHC ($\sqrt{S}=14$ TeV) with subsequent $\tau^{\pm}\to\pi^{\pm}+\nu_{\tau}/\bar{\nu}_{\tau}$
decay. The distribution for a $CP$-even ($CP$-odd) Higgs boson is
shown by the blue dashed line (black dotted line). For a Higgs boson
of $CP$ mixture with $\phi_{\tau}=-\frac{\pi}{4}$ the distribution
is given by the black long-dash dotted line. The distribution for
the Drell-Yan background is shown by the solid red line. }

\label{fig:h_pipi_detcuts} 
\end{figure}

Our method to determine the $CP$-nature of $h$ requires the measurement
of the 4-momenta of the charged prongs $a^{-}$, $a'^{+}$ and their
normalized impact parameter vectors ${\bf \hat{n}}_{\mp}$ in the
laboratory frame. The 4-vectors $n_{\mp}^{\mu}=(0,{\bf \hat{n}}_{\mp})$
are boosted into the $a^{-}a'^{+}$ zero-momentum-frame (ZMF) and
their normalized perpendicular%
\footnote{Perpendicular with respect to the 3-momentum of $a^{-}$ and $a'^{+}$
in the $a^{-}a'^{+}$ ZMF .%
} components ${\hat{n}}_{\perp}^{*\mp}$ define the `unsigned' angle
$\varphi^{*}=\arccos({\bf \hat{n}}_{\perp}^{*+}\cdot{\bf \hat{n}}_{\perp}^{*-})$
with $0\leq\varphi^{*}\leq\pi$. If ${\bf \hat{p}}_{-}^{*}$ is the
normalized $a^{-}$ momentum in the $a^{-}a'^{+}$ ZMF and ${\cal O}_{CP}^{*}={\bf \hat{p}}_{-}^{*}\cdot({\bf \hat{n}}_{\perp}^{*+}\times{\bf \hat{n}}_{\perp}^{*-})$
is a $CP$-odd and $T$-odd triple correlation, the `signed' angle
$\varphi_{CP}^{*}$ with $0\leq\varphi_{CP}^{*}\leq2\pi$ can be defined
by
\begin{equation}
\varphi_{CP}^{*}=\left\{ \begin{array}{ccc}
\varphi^{*} & if & {\cal O}_{CP}^{*}\geq0\,,\\
2\pi-\varphi^{*} & if & {\cal O}_{CP}^{*}<0\,.
\end{array}\right.\label{phistar_CP-1}
\end{equation}
Using this definition, the differential Higgs decay width into a pair
of $\tau$ leptons with respect to $\varphi_{CP}^{*}$ at LO can be
written as \vspace{-2mm}
\begin{equation}
d\Gamma_{h\to\tau^{+}\tau^{-}}\!\approx1-b\left(E_{+}\right)b\left(E_{-}\right)\frac{\pi^{2}}{16}\cos\left(\varphi_{CP}^{*}-2\phi_{\tau}\right).\label{dGamma_dphistar_CP}
\end{equation}
As an example we consider the $h\to\tau^{+}\tau^{-}$ with the $\tau$-decay
mode $\tau^{\mp}\to\pi^{\mp}+\nu_{\tau}/\bar{\nu}_{\tau}$. The spin
analyzing power for this decay mode is $b(E_{\pm})=+1$. The resulting
normalized $\varphi_{CP}^{*}$-distributions for the LHC with $\sqrt{S}=14$
TeV and a Higgs mass of $125$~GeV are shown in Fig.~\ref{fig:h_pipi_detcuts}.
If no cuts are applied on the final state particle momenta as in Fig.~\ref{fig:h_pipi_detcuts},
the distributions are the same for any Higgs production process. 
The dashed blue line shows the distribution for a $CP$-even SM-like
Higgs boson, while the dotted black line shows the distribution for
a $CP$-odd Higgs boson. If the discovered Higgs boson is a $CP$-mixture
the dot-dashed black line shows the distribution for the example of
a mixing angle $\phi_{\tau}=-\frac{\pi}{4}$. The shift of the maximum
of this distribution with respect to the SM-like distribution corresponds
to a shift of $\varphi_{CP}^{*}$ by $2\phi_{\tau}$. If one fits
the appropriate function $u\cdot\cos(\varphi_{CP}^{*}-2\phi_{\tau})+v$
with coefficients $u,v$ to the experimental measurement, one can
extract the mixing angle $\phi_{\tau}$. Similar distributions are
obtained for the other $\tau$-decay channels with a spin analyzing
power smaller than one. The maxima and minima of the distributions
in Fig.~\ref{fig:h_pipi_detcuts} will remain at the current positions.
However, the amplitude of the cosine function will be smaller due
to the prefactor $b\left(E_{+}\right)\, b\left(E_{-}\right)$ in Eq.~\eqref{dGamma_dphistar_CP}.
Furthermore, we have shown in Ref.~\citep{Berge:2014sra} that the
distributions are affected by measurement uncertainties in a non-trivial
way. It is therefore important to calibrate the measurement and derive
reconstruction efficiencies. This can be done by means of the Drell-Yan
background process.

\section{Tau spin correlations for the Drell-Yan production process}

The irreducible and overwhelming background for the $h\to\tau^{-}\tau^{+}$
signal process is the Drell-Yan process $pp\to Z^{*}/\gamma^{*}\to\tau^{-}\tau^{+}$.
Because the masses of the Higgs boson and the $Z$ boson are very
close and because each $\tau$ decay mode includes at least one neutrino,
the two reactions can not be separated by simple cuts. We therefore
investigated~\citep{Berge:2014sra} the $\varphi_{CP}^{*}$ distribution
for the parton level reaction
\begin{equation}
q\,+{\bar{q}}\,\to\gamma^{*},Z^{*}\to\tau^{-}\,+\,\tau^{+}\to a^{-}\,+\, a'^{+}\,+\, X\qquad\label{DYtau0-1}
\end{equation}
at LO and at NLO QCD. Opposite to Higgs production processes, the
matrix element does not factorize  into the $pp\to Z^{*}/\gamma^{*}+X$ production
and $Z^{*}/\gamma^{*}\to \tau^+\tau^-$  decay. Instead the full tau-tau spin density matrix has to
be calculated and appropriately combined with the spin dependent $\tau$-decay
distributions~\eqref{eq:dGamma_dEdOmega}. The differential cross
section at LO, integrated over the polar angles of the $a^{-}$ and
$a'^{+}$ momenta, can be written in the $\tau^{+}\tau^{-}$-ZMF as \vspace{-1mm}
\begin{align}
 & \frac{d\hat{\sigma}_{DY}^{(0)}}{d\phi_{+}d\phi_{-}}\,\sim\,\nonumber \\
 & \qquad1+\frac{\pi^{2}}{32}b(E_{+})b(E_{-})\,\kappa(B_{1},B_{2})\,\cos(\phi_{+}+\phi_{-})\label{dsigmaDY_dphip_dphim}
\end{align}
where $\kappa(B_{1},B_{2})=(a_{\tau}^{B_{1}}a_{\tau}^{B_{2}}-v_{\tau}^{B_{1}}v_{\tau}^{B_{2}})/(a_{\tau}^{B_{1}}a_{\tau}^{B_{2}}+v_{\tau}^{B_{1}}v_{\tau}^{B_{2}})$
and $\phi_{\pm}$ are the azimuthal angles of the $a^{-}$ and $a'^{+}$
momenta in a coordinate system where the direction of the $\tau^{-}$
momentum is chosen to be the $z$-axis, and the momentum of the initial
quark is located in the $x,z$-plane. If one substitutes in Eq.~\eqref{dsigmaDY_dphip_dphim},
e.g. $\phi_{+}$ by the difference of the two azimuthal angles, $\varphi=\phi_{-}-\phi_{+}$
and integrates over $\phi_{-}$ the $\varphi$ dependence drops out.
The corresponding $\varphi_{CP}^{*}$ distribution is therefore flat,
shown by the red line in Fig.~\ref{fig:h_pipi_detcuts}. The $\varphi_{CP}^{*}$
distribution remains flat if NLO QCD corrections are taken into account. 

\begin{figure}[t]
\includegraphics[width=7.8cm]{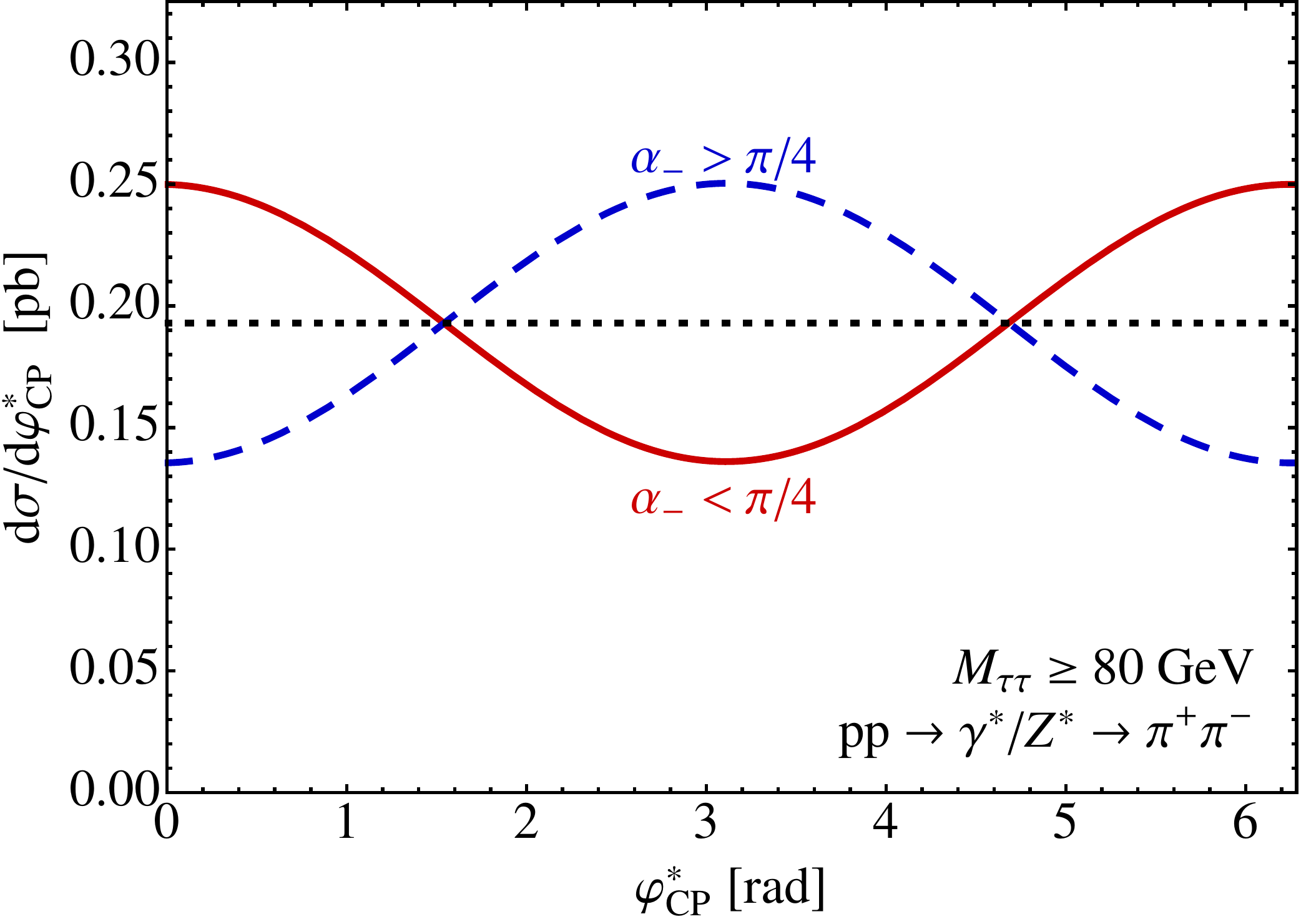}
\caption{Drell-Yan production of $\tau^{-}\tau^{+}$ at the LHC ($\sqrt{S}=14$
TeV, $M_{\tau\tau}\ge80$~GeV, $|\eta_{\pi^{\pm}}|\le1$ ) with subsequent
$\tau^{\pm}\to\pi^{\pm}+\nu_{\tau}/\bar{\nu}_{\tau}$ decay. Events
are split into two categories with $\pi^{-}$ `nearly coplanar' ($\alpha_{-}<\pi/4$,
red solid line) and events with $\pi^{-}$ `nearly perpendicular'
($\alpha_{-}>\pi/4$, dashed blue line) to the $q\tau$ production
plane. The dotted black line indicates the sum of the two distributions
divided by two. }

\label{fig:LHC_Phi_CP_star_nonflat} 
\end{figure}

The signal and background distributions are distorted~\citep{Berge:2014sra}
due to finite experimental resolutions of the charged particle momenta
and the impact parameter measurements. These uncertainties will probably
differ for measurements in the beam direction and perpendicular to
it. We suggest to calibrate this uncertainties using the Drell-Yan
background process in the following way: Eq.~\eqref{dsigmaDY_dphip_dphim}
shows that a non-uniform distribution is obtained if neither $\phi_{-}$
nor $\phi_{+}$ are integrated out. Instead we divide the events into
two classes: one class which collects events where the $\pi^{-}$
momentum is preferably parallel to the production plane, the second
class collects events where the $\pi^{-}$ momentum is preferably
perpendicular to this plane. The production plane for the scattering
reaction Eq.~\eqref{DYtau0-1} is defined by the beam axis and the
$\tau^{-}$ momentum in the laboratory frame. Our method to extract
the mixing angle $\phi_{\tau}$ was, however, specifically defined
such that no reconstruction of the $\tau$-momenta needs to be performed.
Therefore, we suggest to use the alternative definition in terms of
measurable quantities:\vspace{-1mm}
\begin{eqnarray}
\cos\alpha_{-} & = & \left|\frac{{\bf \hat{e}_{z}}\times{\bf \hat{p}}_{L-}}{\left|{\bf \hat{e}_{z}}\times{\bf \hat{p}}_{L-}\right|}\cdot\frac{{\bf \hat{n}_{-}}\times{\bf \hat{p}}_{L-}}{\left|{\bf \hat{n}_{-}}\times{\bf \hat{p}}_{L-}\right|}\right|\,\,.\label{cosalphabeta}
\end{eqnarray}
Here ${\bf \hat{p}}_{L-}$ is the $\pi^{-}$ direction of flight in
the laboratory frame, ${\bf \hat{e}}_{z}$ points along the direction
of one of the proton beams and ${\bf \hat{n}_{-}}$ is the impact
parameter vector of the $\pi^{-}$ in the laboratory frame. With the
definition~\eqref{cosalphabeta} we define events with $\pi^{-}$
being `nearly coplanar' (`nearly perpendicular') by requiring ${\alpha}_{-}<\pi/4$
(${\alpha}_{-}>\pi/4$). The resulting distributions for Drell-Yan
production at the LHC with $\sqrt{S}=14$ TeV and the direct decays
$\tau^{\mp}\to\pi^{\mp}+\nu_{\tau}/\bar{\nu}_{\tau}$ are shown in
Fig.~\eqref{fig:LHC_Phi_CP_star_nonflat}. 
The solid red line shows the distribution for ${\alpha}_{-}<\pi/4$
and the dashed blue line the distribution for ${\alpha}_{-}>\pi/4$.
The maximum of the distribution for ${\alpha}_{-}>\pi/4$ and the
minimum for ${\alpha}_{-}<\pi/4$ are located at $\varphi_{CP}^{*}=\pi$.
The form of the curves in Fig.~\eqref{fig:LHC_Phi_CP_star_nonflat}
are essentially due to $Z^{*}$ boson exchange. The resonant $Z^{*}$
boson peak has been included ($M_{\tau\tau}\ge80$~GeV) to increase
the amplitude. For larger $M_{\tau\tau}$-cuts, the amplitude becomes
smaller because the $\gamma^{*}$ contribution increases with respect
to the $Z^{*}$ contribution. If one assumes pure $\gamma^{*}$ exchange
the solid red and dashed blue curves in Fig.~\eqref{fig:LHC_Phi_CP_star_nonflat}
would be reversed. This is because the coefficient for $Z^{*}$ boson
exchange in Eq.~\eqref{dsigmaDY_dphip_dphim} is $\kappa(Z,Z)\approx0.98$,
while for photon exchange it is $\kappa(\gamma,\gamma)=-1$. Furthermore,
the amplitude of the two distributions also depends on the considered
rapidities of the two final state pions. Here, the rapidity cuts $|\eta_{\pi^{\pm}}|\le1$
have been applied to enhance the difference between the two curves.
As in the case of $\tau$-pair production via Higgs boson exchange,
the distributions depend on experimental uncertainties and reconstruction
efficiencies. Our numerical simulation shows that both signal and
background reactions are affected in a similar way by these uncertainties.
The background distributions with its large numbers of events can
therefore be used to calibrate the $h\to\tau^{+}\tau^{-}$ signal
reaction.\\
\vspace{-7mm}

\section{Concluding Remarks}

We have proposed a method to probe for a possible Higgs sector $CP$-violation
in the $h\to\tau^{+}\tau^{-}$ decay at the LHC. The corresponding
mixing angle $\phi_{\tau}$ can be extracted by measuring the spin
correlations of the $\tau$-lepton pairs. These spin correlations
manifest themselves in characteristic distributions of the final $\tau$
decay products. Our method uses the momentum directions and the impact
parameters of the final charged prongs of the $\tau$ decay to reconstruct
these $\tau$-spin correlations. We furthermore investigated the largest
background to the Higgs boson signal, the Drell-Yan production process.
By dividing these background events into two subsets, we proposed
a method to calibrate measurement uncertainties of the signal process.
Using these results we performed~\citep{Berge:2014sra} an estimate
of the precision of the mixing angle $\phi_{\tau}$ which might be
achievable at the LHC. For this evaluation we included all major $\tau$-decay
channels and simulated the $\varphi_{CP}^{*}$ distribution by a Monte
Carlo program including resolution estimates for the momenta and impact
parameters of the final charged prongs~\citep{Berge:2014sra}. We
furthermore assumed signal to background ratios taken from Ref.~\citep{ATLtauconf}.
For the LHC with $\sqrt{S}=14$~TeV we performed simulations for
$150\, fb^{-1}$, $500\, fb^{-1}$ and the high-luminosity LHC upgrade
goal of $3\, ab^{-1}$. The precision of $\phi_{\tau}$ for these
luminosities is estimated to $27^{\circ}$, $14.3^{\circ}$, and $5.1^{\circ}$
respectively.

\section*{Acknowledgments}

We wish to thank the members of the $M_{\tau\tau}$ working group
of the Helmholtz Alliance ``Physics at the Terascale'' for discussions.
The work of W.B. is supported by B.M.B.F., contract 05H12PAE, and
S.K. is supported by Deutsche Forschungsgemeinschaft through Graduiertenkolleg
GRK 1675.




 \bibliographystyle{elsarticle-num}
\bibliography{bibliography_BBK}







\end{document}